\documentclass[twocolumn,english,prl,superscriptaddress,amsmath,amsfonts,twocolumn]{revtex4}
\usepackage[T1]{fontenc}
\usepackage[utf8]{inputenc}
\usepackage{relsize}
\usepackage{graphicx}

\makeatletter

\DeclareRobustCommand{\lyxmathsym}[1]{\ifmmode\begingroup\def\b@ld{bold}
  \def\rmorbf##1{\ifx\math@version\b@ld\textbf{##1}\else\textrm{##1}\fi}
  \mathchoice{\hbox{\rmorbf{#1}}}{\hbox{\rmorbf{#1}}}
  {\hbox{\smaller[2]\rmorbf{#1}}}{\hbox{\smaller[3]\rmorbf{#1}}}
  \endgroup\else#1\fi}

\makeatother

\usepackage{babel}

\begin{document}

\title{Quantum and classical correlations in waveguide lattices}

\author{Yaron Bromberg}

\affiliation{Department of Physics of Complex Systems, Weizmann Institute of Science,
Rehovot 76100, Israel}

\author{Yoav Lahini}

\affiliation{Department of Physics of Complex Systems, Weizmann Institute of Science,
Rehovot 76100, Israel}

\author{Roberto Morandotti}

\affiliation{Institute National de la Recherche Scientifique, Varennes, Qúbec,
Canada}

\author{Yaron Silberberg}

\email[]{yaron.silberberg@weizmann.ac.il}

\affiliation{Department of Physics of Complex Systems, Weizmann Institute of Science,
Rehovot 76100, Israel}

\date{March 31, 2009}
\begin{abstract}
We study quantum and classical Hanbury Brown-Twiss correlations in
waveguide lattices. We develop a theory for the propagation of
photon pairs in the lattice, predicting the emergence of nontrivial
quantum interferences unique to lattice systems. Experimentally, we
observe the classical counterpart of these interferences using
intensity correlation measurements. We discuss the correspondence
between the classical and quantum correlations, and consider
path-entangled input states which do not have a classical analogue.
Our results demonstrate that waveguide lattices can be used as a
robust and highly controllable tool for manipulating quantum states,
and offer new ways of studying the quantum properties of light.
\end{abstract}
\maketitle
Lattices of coupled waveguides have been shown to be extremely versatile
in manipulating the flow of light \cite{RevNature,ReviewRep}. Such
systems enabled direct observation of optical analogues of many fundamental
quantum mechanical effects such as Bloch oscillations \cite{Bloch1,Bloch2},
Anderson localization \cite{And1,segev,lahini}, quantum Zeno effect
\cite{Zeno}, quantum walks \cite{QRW} and many others \cite{RevNature,ReviewRep,Zener}.
However, these effects were all related to the wave properties of
light rather than to its particle nature. To observe the quantum properties
of light, one should consider correlations between single photons.
Here we show that photon pairs propagating in waveguide lattices develop
nontrivial quantum correlations unique to lattice systems. We experimentally
observe some of these features, albeit with reduced contrast, using
Hanbury Brown-Twiss intensity correlation measurements with phase-averaged
coherent states. Our results suggest that waveguide lattices can be
used as a robust and highly controllable tool for manipulating quantum
states in an integrated manner, offering new ways of studying quantum
properties of light in structured media.

Consider a lattice that is made of identical optical waveguides, each
supporting a single transverse mode, as shown schematically in Fig.
1a. The evolution of the quantized electromagnetic field in each waveguide
is given by the Heisenberg equation for the bosonic creation operator
$a^{\dagger}$. For a lattice with nearest-neighbors couplings, the
Heisenberg equation for the creation operator of the $k^{th}$ waveguide
is given by:
\begin{equation}
i\frac{n}{c}\frac{\partial a_{k}^{\dagger}}{\partial t}=
i\frac{\partial a_{k}^{\dagger}}{\partial z}=\beta
a_{k}^{\dagger}+C_{k,k+1}a_{k+1}^{\dagger}+
C_{k,k-1}a_{k-1}^{\dagger},\label{eq:eg1}
\end{equation}

where $z$ is the spatial coordinate along the propagations axis,
$\beta$ is the propagation constant of the waveguides, $C_{k,k\pm1}$
are the coupling constants between adjacent waveguides, and $c/n$ is
the speed of light in the medium. The creation operators at any
point along the propagation are calculated by integrating
Eq.~(\ref{eq:eg1}):
\begin{equation} a_{k}^{\dagger}(z)=e^{i\beta
z}\sum_{l}U_{k,l}(z)a_{l}^{\dagger}(z=0),\:
U_{k,l}(z)=\left(e^{i\beta zC}\right)_{k,l}.\label{eq:2}
\end{equation}

$U_{k,l}(z)$ is a unitary transformation given by calculating the
exponent of the coupling matrix $izC_{k,l}$, which describes the
amplitude for the transition of a single photon from waveguide $l$
to waveguide $k$. Since any input state can be expressed with the
creation operators $a_{l}^{\dagger}$ and the vacuum state
$\left|0\right\rangle $, the evolution of non classical states along
the lattice can be calculated using Eq.~(\ref{eq:2}). When a single
photon is coupled to waveguide $l$, the input state
$a_{l}^{\dagger}\left|0\right\rangle \equiv\left|1\right\rangle
_{l}$ will evolve to the superposition
$\sum_{k}U_{l,k}^{*}\left|1\right\rangle _{k}$, where
$\left|1\right\rangle _{k}$ is the state of a single photon
occupying waveguide $k$. However, measurements of the probability
distribution of single photons are not enough to reveal the quantum
properties of light, as the probability distribution of a single
photon $|U_{k,l}(z)|^{2}$ evolves in the same way as the intensity
distribution of classical light \cite{QRW,Kim04,Agarwal07}. The
quantum mechanical properties of light are observed when
correlations between the propagating photons are considered. In this
Letter we focus on the evolution of the photon-number correlation
function $\Gamma_{q,r}=\left\langle
a_{q}^{\dagger}a_{r}^{\dagger}a_{r}a_{q}\right\rangle $, when two
indistinguishable photons are injected into the lattice.

We start by analyzing the simplest example of two coupled
waveguides. The coupling matrix in this case is
$\overleftrightarrow{C}= \left(\begin{array}{cc} 0 & C\\ C &
0\end{array}\right)$, and the transformation $U_{k,l}(z)$ is
therefore $\overleftrightarrow{U}(z)=\left(\begin{array}{cc} cos(Cz)
& isin(Cz)\\ isin(Cz) & cos(Cz)\end{array}\right)$. The coupler acts
as a beam splitter, with the reflection and transmission
coefficients varying continuously along the propagation. If two
photons are injected to the coupler, one to each waveguide, the
average photon number at each of the waveguides is constant since
$n_{1(2)}(z)=\left\langle a_{1(2)}^{\dagger}a_{1(2)}\right\rangle
=|U_{11}|^{2}+|U_{12}|^{2}=1$. The nonclassical nature of the light
is revealed by considering $\Gamma_{1,2}$, the probability to detect
exactly one photon at each waveguide (a coincidence measurement).
Using Eq.~(\ref{eq:2}) we obtain
$\Gamma_{1,2}(z)=|U_{11}U_{22}+U_{12}U_{21}|^{2}=Cos^{2}(2Cz)$.
Since two paths lead to the final state of one photon at each
waveguide, they interfere and the probability for a coincidence
measurement oscillates along the propagation. After propagating
exactly half a coupling length $z_{BS}=\pi/4C$, we find
$\Gamma_{1,2}(z_{BS})=0$. At this point the two photons bunch, and
are found together in either one of the two waveguides.
$\overleftrightarrow{U}(z_{BS})$ is identical to the transformation
of a symmetric beam splitter, and the coincidence measurement
vanishes in the same manner as in the Hong-Ou-Mandel (HOM)
interferometer \cite{HOM}. Since a pair of coupled waveguides is
equivalent to a beam splitter, it is possible to cascade several of
them in order to implement quantum gates in an integrated manner
\cite{Obrien}. A lattice of many coupled waveguides enriches the
variety of correlations obtained in integrated structures, as we
show bellow.

\begin{figure}
\includegraphics[clip,width=1\columnwidth]{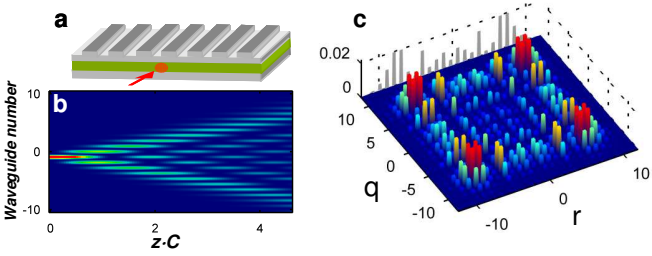}

\caption{(a) A schematic view of the waveguide lattice used in the
experiment. The red arrow represents the input light beam. (b) The
calculated probability distribution $\left\langle
n_{q}(z)\right\rangle $ of a single photon injected to the central
waveguide of a periodic lattice, as a function of the propagation
distance. The photon couples coherently from each waveguide to its
neighbors, and the probability distribution concentrates at two
outer lobes. (c) The calculated correlation matrices $\Gamma_{q,r}$,
representing the probability to detect at the output of the lattice
exactly one photon at waveguide $r$ and one photon at waveguide $q$,
when both photons are coupled to a single waveguide at the center of
the lattice, i.e. $\left|\varphi_{0}\right\rangle
\equiv1/\sqrt{2}a_{0}^{\dagger2}\left|0\right\rangle $. This matrix
is a simple product of two single-photon distributions, thus showing
no quantum interference. The grey bars are obtained by summing the
matrix along one axis, representing the results of a single photon
measurement. }

\end{figure}
\begin{figure}
\includegraphics[clip,width=1\columnwidth]{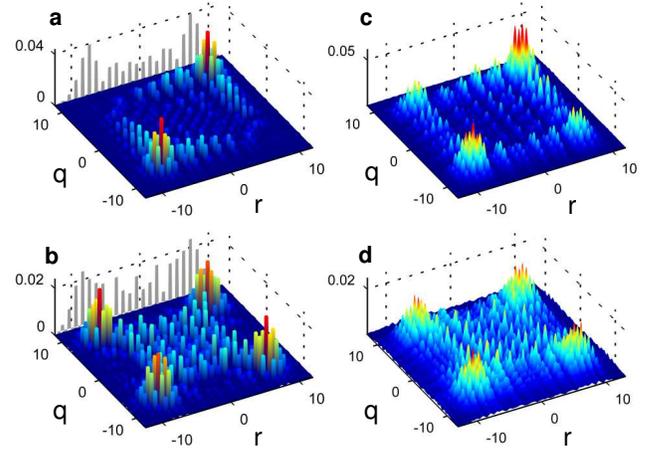}
\caption{Quantum and classical correlations in waveguide lattices.
(a) The correlation matrix $\Gamma_{q,r}$ when the photons are
coupled to two adjacent waveguides i.e. $\left|\varphi_{1}\right
\rangle \equiv a_{0}^{\dagger}a_{1}^{\dagger}\left|0\right\rangle $.
The two photons exhibit bunching, and will emerge from the same side
of the lattice. (b) The correlation matrix when the two photons are
coupled to two waveguides separated by one waveguide,
$\left|\varphi_{2}\right\rangle \equiv
a_{-1}^{\dagger}a_{1}^{\dagger}\left|0\right\rangle $ . Here the two
photons will emerge either both from the lobes, or both from the
center. (c) Measured classical intensity correlations
$\Gamma_{q,r}^{(c)}$, corresponding to the
$\left|\varphi_{1}\right\rangle $ input state. (d) Measured
classical intensity correlations $\Gamma_{q,r}^{(c)}$, corresponding
to the $\left|\varphi_{2}\right\rangle $ input state.}
\end{figure}

We now turn to study the quantum properties of a periodic lattice
with a large number of identical waveguides, where all the coupling
constants are equal $C_{n,n\pm1}=C$. As long as the photons are far
from the boundaries of the lattice, Eq.~(\ref{eq:2}) yields
$U_{q,k}(z)=i^{q-k}J_{q-k}(2Cz)$, where $J_{q}$ is the $q^{th}$
Bessel function \cite{ReviewRep,Yariv}. When a single photon is
coupled to waveguide $k$, it will evolve to waveguide $q$ with a
probability $n_{q}=|U_{q,k}(z)|^{2}=J_{q-k}(2Cz)^{2}$. The photon
spreads across the lattice by coupling from one waveguide to its
neighbors in a pattern characterized by two strong 'ballistic'
lobes, as shown in Fig. 1b. If a second photon is coupled to another
waveguide $l$, then the average photon number at waveguide $q$ is
simply the incoherent sum
$n_{q}=<a_{q}^{\dagger}a_{q}>=J_{q-k}(2Cz)^{2}+J_{q-l}(2Cz)^{2}$.
Once again, the quantum nature of light is revealed by considering
the correlations between the two photons. In the following we study
the correlation $\Gamma_{q,r}$ for three distinct two-photon input
states: \textit{i)} Both photons are coupled to a single waveguide
at the center of the lattice, i.e. $\left|\varphi_{0}\right\rangle
\equiv1/\sqrt{2}a_{0}^{\dagger2}\left|0\right\rangle $ \textit{(ii)}
The two photons are coupled to two adjacent waveguides
$\left|\varphi_{1}\right\rangle \equiv
a_{0}^{\dagger}a_{1}^{\dagger}\left|0\right\rangle $ and
\textit{iii)} The two photons are coupled to two waveguides,
separated by one waveguide, $\left|\varphi_{2}\right\rangle \equiv
a_{-1}^{\dagger}a_{1}^{\dagger}\left|0\right\rangle $. The
correlation matrix $\Gamma_{q,r}$ represents the probability to
detect one photon at waveguide $q$ and its twin photon at waveguide
$r\neq q$. The probability to detect both photons at the same
waveguide $q$ is given by $\Gamma_{q,q}/2$.

Fig.~1c depicts the correlation matrix $\Gamma_{q,r}$ at the output
of the lattice, when the two photons are coupled to the same input
waveguide (the $\left|\varphi_{0}\right\rangle $ input state). In
this case, there is no interference and the correlation matrix is
just a product of the two classical probability distributions,
$\Gamma_{q,r}=2|U_{q0}U_{r0}|^{2}$. The correlation map is
characterized by four strong lobes at the corners of the matrix,
resulting from the tendency of the photons to propagate in the
ballistic directions.

When the two photons are coupled to two neighboring sites, i.e. the
$\left|\varphi_{1}\right\rangle $ state, the correlation map changes
considerably as shown in Fig.~2a. The most obvious feature is the
vanishing of the two 'off-diagonal' lobes: the photons tend to bunch
to the same lobe. This can be thought of as a generalized HOM
interference. Two paths lead to a coincidence measurement between
waveguide $q$ and waveguide $r$: either the photon from waveguide
$0$ propagates to waveguide $q$ and the photon from waveguide $1$ to
waveguide $r$, or vice versa - from waveguide $0$ to waveguide $r$
and from waveguide $1$ to waveguide $q$ \cite{Zeilinger95}. These
paths are complex and involve hopping of the photons between many
waveguides, nevertheless they interfere and the correlation matrix
is thus given by $\Gamma_{q,r}=|U_{q0}U_{r1}+U_{q1}U_{r0}|^{2}$. The
destructive interference which leads to vanishing of the
off-diagonal lobes can be mathematically traced to the inherent
$90\lyxmathsym{º}$ phase shift associated with nearest-neighbors
coupling. We note that since the photons tunnel between the
waveguides continuously, the visibility of the quantum interference
only weakly depends on the overall length of the lattice.

The four lobes are recovered when the photons are initiated in the
$\left|\varphi_{2}\right\rangle $ state, i.e. with one waveguide
separation between the input sites (Fig 2b). However, this state
also contains strong non-classical features - note the differences
between Fig. 1c and 2b. The photon pair exhibit bunching but with a
different symmetry: if one photon is detected in between the lobes,
the probability to detect the second photon in a lobe vanishes, even
though a single photon is most likely to reach the lobes. Similarly,
if one photon is detected in a lobe, it is certain that the other
photon is also in a lobe. It is important to note that in the above
examples, the quantum interference emerges since the two photons are
indistinguishable. Interferences with distinguishable particles are
possible, but require an entangled input state \cite{Omar06}.

Some of the special features of these quantum mechanical
correlations can be captured using intensity correlation
measurements with classical light \cite{PaulReview,Boyd04,Resch}.
The quantum mechanical probability to detect one photon at waveguide
$q$ and its twin photon at waveguide $r$ is related to the classical
intensity correlation function $\Gamma_{q,r}^{(c)}=\left\langle
I_{q}I_{r}\right\rangle $ where $\left\langle \cdot\right\rangle $
denotes statistical (or temporal) averaging. This non-local
intensity correlation function is usually discussed in the context
of the classical Hanbury Brown-Twiss effect and its quantum
interpretation \cite{HBT,Glauber63}. We studied experimentally
intensity correlations at the output of a periodic waveguide
lattice. The lattice of $89$ identical waveguides was fabricated on
an AlGaAs substrate using e-beam lithography, followed by reactive
ion etching. Each waveguide is $8mm$ long and $4\mu m$ wide
\cite{Hagai98}. The tunneling parameter between sites $C$ is
determined by the etch depth of the sample ($1.3\mu m$) and by the
distance between neighboring waveguides ($4\mu m$) and was measured
to be $290m^{-1}$. We used an OPO (Spectra-Physics, OPAL) pumped by
a mode-locked Ti:Sapphire laser (Spectra-Physics Tsunami) to
generate $150fs$ pulses, at a wavelength of $1530nm$ with $80MHz$
repetition rate. The average power was on the order of $0.1mW$, thus
nonlinear effects were negligible.

Two-photon input states can be mimicked by injecting into the
lattice two phase-averaged coherent states: two coherent states with
the same mean photon number and a fluctuating relative phase
\cite{PaulReview,Mandel83}. Thus, we injected two coherent beams
into two different sites of the lattice \cite{szameit07}, and
randomized their relative phase with a spatial light modulator. For
each phase realization, the intensity profile at the output facet of
the slab was imaged on an infrared camera, and the intensity
correlations between all the waveguides were computed. The
intensity-correlation function $\Gamma_{q,r}^{(c)}$ was obtained by
averaging the measured correlations over many phase realizations.
The measured intensity correlations for the nearest neighbors
(mimicking the $\left|\varphi_{1}\right\rangle $ state) and
next-nearest neighbors ($\left|\varphi_{2}\right\rangle $ state)
inputs are presented in figures 2c and 2d, correspondingly. The
patterns are strikingly similar to the corresponding quantum
correlations $\Gamma_{q,r}$, except for the reduced contrast.
Indeed, for two incoherent sources coupled to waveguides $k$ and
$l\neq k$, the classical correlation is given by
$\Gamma_{q,r}^{(c)}(z)=I_{0}^{2}
\left(|U_{ql}U_{rk}+U_{qk}U_{rl}|^{2}+
|U_{ql}U_{rl}|^{2}+|U_{qk}U_{rk}|^{2}\right)$, where $I_{0}$ is the
intensity coupled to each waveguide. The last two terms are
responsible for the reduced contrast. Thus, while the 'off-diagonal'
lobes practically vanish for the quantum input state
$\left|\varphi_{1}\right\rangle $, a straightforward calculation
shows that for classical light
$\Gamma_{q,r}^{(c)}>\frac{1}{3}\sqrt{\Gamma_{q,q}^{(c)}\Gamma_{r,r}^{(c)}}$,
in agreement with our experimental results. Still these classical
Hanbury Brown-Twiss type intensity correlations do echo many of the
special features of the quantum correlations of Fig. 2a,b.

\begin{figure}
\includegraphics[clip,width=1\columnwidth]{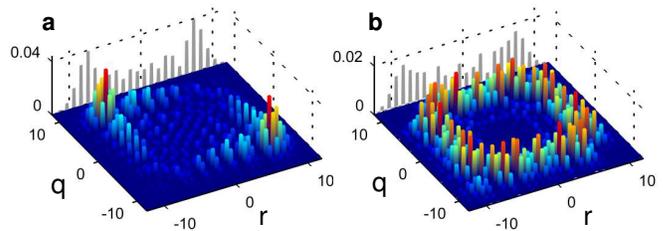}
\caption{Quantum correlation maps $\Gamma_{q,r}$ for path-entangled
input states. (a) The two photons are injected together to either of
two neighboring waveguides, $\left|\psi^{(+)}\right\rangle
=\frac{1}{2}\left(a_{1}^{\dagger2}+a_{0}^{\dagger2}\right)\left|0\right\rangle
$. The correlation is significant only in the 'off-diagonal' peaks,
which indicates that the two photons will emerge from opposite sides
of the lattice. (b) The correlation map for an input state with two
photons in either of two waveguides with one waveguide separation,
where there is a $\pi$-phase between the two possibilities,
$\left|\psi^{(-)}\right\rangle
=\frac{1}{2}\left(a_{1}^{\dagger2}-a_{-1}^{\dagger2}\right)\left|0\right\rangle
$. }
\end{figure}

The great potential of waveguide lattices for quantum information
probably lies in the extension of these concepts to non-uniform
lattices in order to specially design the correlation properties,
and in the utilization of these multiport systems for handling more
complex quantum states. As an example for the latter, consider the
propagation of a path-entangled input state with two photons in
either of two neighboring waveguides, $\left|\psi^{(+)}\right\rangle
=\frac{1}{2}\left(a_{1}^{\dagger2}+a_{0}^{\dagger2}\right)\left|0\right\rangle
$. The calculated correlation map for this case is presented in Fig
3a. The correlation in the 'diagonal' peaks completely vanishes and
is significant only in the 'off-diagonal' peaks. Accordingly, the
two photons will always separate and emerge from different sides of
the lattice. The corresponding correlation map violates the
Cauchy-Schwarz inequality
$\Gamma_{q,r}<\sqrt{\Gamma_{q,q}\Gamma_{r,r}}$, indicating that this
case has no apparent classical analog. As another illustration,
consider the state $\left|\psi^{(-)}\right\rangle
=\frac{1}{2}\left(a_{1}^{\dagger2}-a_{-1}^{\dagger2}\right)\left|0\right\rangle
$. Now the photons are in either of two next-nearest neighboring
waveguides, yet with a $\pi$-phase shift. The correlation map shown
in Fig. 3b reveals that in this case, one photon will always reach a
lobe while the other will always reach the center.

In this Letter we studied the evolution of photon pairs along
periodic lattices, and have shown that the resulting quantum
correlations strongly depend on the input states. We compared our
predictions with a classical wave theory, and experimentally
demonstrated that some features can be obtained using classical
intensity correlation measurements, yet with reduced contrast. The
correspondence between the classical and quantum nature of light can
be further studied by considering the evolution of quantum
correlations in the presence of dephasing, which can be introduced
via lattice inhomogeneities. Furthermore, waveguide lattices offer
new possibilities as they can be designed in ways that are not
feasible using bulk or fiber optical systems. It will be especially
interesting to study the effect of such lattices on the propagation
of other types of non-classical light, such as squeezed states
\cite{Agarwal08} and cat states.

We would like to thank H. S. Eisenberg, N. Bar-Gill and H. B. Perets
for valuable help. Financial support by NSERC and CIPI (Canada), and
EPRSC (UK) is gratefully acknowledged. YL is supported by the Adams
Fellowship of the Israel Academy of Sciences and Humanities.

\end{document}